# Astro2020 Science White Paper

# Imaging Cool Giant Planets in Reflected Light: Science Investigations and Synergy with Habitable Planets

**Principal Author:**
Name: Mark Marley
Institution: NASA Ames Research Center
Email: Mark.S.Marley@NASA.gov
Phone: (650) 604-0805

**Co-authors:**
Nikole Lewis (Cornell Univ.) and Giada Arney (GSFC), Vanessa Bailey (JPL), Natasha Batalha (UCSC), Charles Beichman (NESI), Björn Benneke (U. Montreal), Jasmina Blecic (NYUAD), Kerri Cahoy (MIT), Jeffrey Chilcote (U. Notre Dame), Shawn Domagal-Goldman (GSFC), Courtney Dressing (UC Berkeley), Michael Fitzgerald (UCLA), Jonathan Fortney (UCSC), Richard Freedman (ARC), Dawn Gelino (NESI), John Gizis (U. Deleware), Olivier Guyon (U. Arizona), Thomas Greene (ARC), Heidi Hammel (AURA), Yasuhiro Hasegawa (JPL), Nemanja Jovanovic (Caltech), Quinn Konopacky (UCSD), Ravi Kopparapu (GSFC), Michael Liu (U. Hawaii), Eric Lopez (GSFC), Jonathan Lunine (Cornell), Roxana Lupu (BAER), Bruce Macintosh (Stanford), Kathleen Mandt (JHU), Christian Marois (U. Victoria), Dimitri Mawet (Caltech), Laura Mayorga (CfA), Caroline Morley (U.T. Austin), Eric Nielsen (Stanford), Aki Roberge (GSFC), Eugene Serabyn (JPL), Andrew Skemer (UCSC), Karl Stapelfeldt (JPL), Channon Visscher (Doredt Univ.), Jason Wang (Caltech)

**Abstract:**
Planned astronomical observatories of the 2020s will be capable of obtaining reflected light photometry and spectroscopy of cool extrasolar giant planets. Here we explain that such data are valuable both for understanding the origin and evolution of giant planets as a whole and for preparing for the interpretation of similar datasets from potentially habitable extrasolar terrestrial planets in the decades to follow.

## Introduction

WFIRST and the extremely large telescopes with specialized instruments, such as TMT/PSI or GMT/GMagAO-X, planned for the next decade will have the capability to detect and characterize giant planets in reflected light. Here we argue that scattered light spectroscopy of giant planets offers outstanding science opportunities and should be considered an important component of the exoplanet science portfolio in the next decade for two reasons: (1) such data will provide scientifically valuable new insights into extrasolar planets; and (2) these planets provide a critical training ground for developing and testing methods for interpreting scattered light spectra of extrasolar planets in preparation for LUVOIR, HabEx, or the WFIRST + Starshade mission.

## Exoplanet Scattered Light Spectroscopy

Modeling and interpretation of directly-imaged planets in scattered light is intrinsically different from the analysis of transiting planets that has comprised much of exoplanet science to this point. Incident rays detected in transit spectra pass through an atmosphere at fairly low pressures where, once scattered, there is generally a very low chance of photons being scattered back into the beam which reaches the observer (see Robinson 2017). This somewhat forgiving set of circumstances has helped fuel the boom in exoplanet transit spectroscopy.

The interpretation of photons received from the partially illuminated disk of a planet as observed by direct imaging is more complex. Reflected light observations are sensitive to a multitude of scattering pathways through the atmosphere. Photons arriving at the limb are often primarily singly scattered from high-altitude hazes, usually produced by photochemistry, while photons returning from the center of the disk have probed much deeper, by factors of 100 or more in pressure, into the atmosphere, and encountered multiple cloud layers (e.g., Batalha et al. 2019). Necessarily, the abundance of atmospheric gases of interest (e.g., $CH_4$ or $H_2O$) must be derived along with the entire atmospheric haze, cloud, and thermal structure in order to interpret the reflected light spectrum of a planet. This is because the measured depths of gaseous absorption bands depend upon both the continuum flux level, set by cloud or gas scattering, and the column of absorbing gas above the cloud layer, which in turn depends on the thermal profile and gravity of the planet. Unlike for the case of transiting planets, which usually have measured mass and radius – and thus gravity and scale height – as prior constraints, there will be few to no constraints on radius for directly-imaged planets, compounding the challenge (GAIA should provide masses for many known RV planets). Given such challenges there is a need for proving grounds where observing strategies and analysis algorithms can be tested on easier targets than potentially habitable planets.

## Cool Giant Planet Science

The first extrasolar planet to be detected around a main sequence star was a hot Jupiter, 51 Peg b, with an equilibrium temperature, $T_{eq}$, of over 1200 K. However, despite the subsequent stunning successes of this field, essentially all transiting planets since characterized have $T_{eq}$ > 500 K. The observed atmosphere in these cases is strongly impacted by incident flux and there is a deep radiative zone which acts to decouple the dynamics and chemistry of the observed atmosphere from the deeper interior. Only for cooler, more distant giant planets—which have never been probed by transit methods—do we expect atmospheric thermal profiles to be more like our own Solar System giant planets, with an observable atmosphere directly coupled to the deep interior.

Because of this distinction between the hot Jupiters and cooler giants, we do not know if various trends identified among the transiting planets hold for all giants. For example much effort has focused on determining if there is a mass-metallicity trend in hot Jupiter atmospheres that is comparable to that seen in solar system giants (where metallicity falls with increasing mass, e.g., Wakeford et al. 2018). For giant planets somewhat warmer than our own Jupiter, $H_2O$, $CH_4$, and $NH_3$ will all be present in gaseous form, allowing measurement of C, N, and O abundances as a function of planetary mass and orbit, stellar metallicity, and other parameters. Only by directly observing cool, distant extrasolar giants will a truly direct comparison be possible.

Furthermore, because of the role cool giants planets may play in delivering volatiles to potentially habitable planets (e.g., Marov & Ipatov 2018), the record of this process preserved in their own atmospheric abundances, and the great diversity of atmospheric chemistry, cloud processes, photochemistry, and dynamics likely present in their atmospheres, the cool giants (with masses ranging from 'sub-Neptune' to 'super-Jupiter') are worthy of detailed characterization.

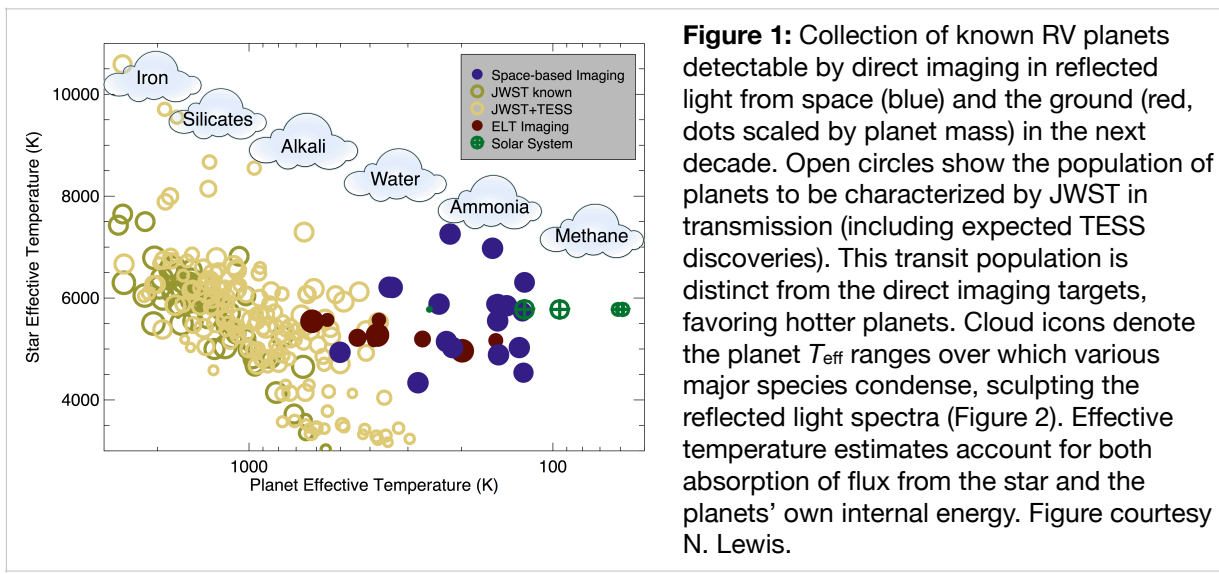

**Figure 1:** Collection of known RV planets detectable by direct imaging in reflected light from space (blue) and the ground (red, dots scaled by planet mass) in the next decade. Open circles show the population of planets to be characterized by JWST in transmission (including expected TESS discoveries). This transit population is distinct from the direct imaging targets, favoring hotter planets. Cloud icons denote the planet $T_{eff}$ ranges over which various major species condense, sculpting the reflected light spectra (Figure 2). Effective temperature estimates account for both absorption of flux from the star and the planets' own internal energy. Figure courtesy N. Lewis.

There are many long orbital period giants from radial velocity searches which are excellent targets for space-based coronagraphs (Figure 1) and more will be found by future direct imaging surveys. Some characterization of these planets may occur serendipitously by missions focusing on habitable planet discovery and characterization. However only purposeful surveys will collect sufficient data to permit an understanding of the comparative planetology of cool giant planets. We urge that any future direct imaging missions recognize the importance of cool giant planet science and plan accordingly for such science to be accommodated, e.g., through Design Reference Missions, instrument design, selection of participating scientists, operational development, and so on. Without mindful instrument design and operation choices, giant exoplanet science could be precluded.

**Giants as a Training Ground**

In addition to their intrinsic science value, giants are independently useful as calibration and training targets for the effort to characterize potentially habitable planets. Specifically, the complex multiple-scattering atmosphere codes required to properly interpret the reflected-light datasets that will be obtained by future direct imaging missions generally do not exist within the exoplanet community today. These must be constructed and tested without the common simplifications that are appropriate for the analysis of exoplanet transit spectra. Reflected-light spectra of potentially habitable terrestrial planets will almost certainly have low signal-to-noise ratios (SNR), contain elements of cloud and photochemical haze scattering as well as gaseous absorption, and consequently be challenging to interpret. The community needs assurance that the interpretation methods and models used to study the data are robust. Likewise the expertise among astronomers necessary to plan and interpret reflected-light observations must be nurtured. Most exoplanet astronomers today have little to no experience with such datasets.

Fortunately, giant-planet reflected-light spectra will offer an excellent, high SNR training ground to develop observational and theoretical methods to interpret such reflected-light data for planets (Marley et al. 1999; Sudarsky et al. 2003, 2005; Cahoy et al. 2010). Bulk composition, equilibrium chemistry, cloud composition, and photochemical pathways are relatively well understood for giant planets. Thus, forward models and retrieval methods face fewer unknowns than they do for terrestrial planets (Venus, Earth, and Mars exhibit far greater atmospheric diversity than Jupiter, Saturn, and Neptune, for example) and are natural testing grounds for observational interpretation methods such as spectroscopic retrievals. They are also a control sample to confirm that habitability false positives do not unexpectedly appear, perhaps from unanticipated photochemical pathways for example. Giant planets can therefore serve as high quality targets for the interpretation of reflected light data before these same techniques are applied to terrestrial planets.

**Giant Planet Science Investigations**

Below we briefly highlight three areas where giants provide interesting intrinsic science and can enhance direct imaging habitable planet science.

*Composition*

The atmospheres of cool gas giant planets are predominantly $H_2$-He with a few percent sprinkling of other notable gasses, predominantly $CH_4$, $NH_3$, $H_2O$, and – for warm planets – detectable Na and K. These gases and their condensates sculpt the optical reflected-light spectra. What is of interest to observers and theorists alike is the abundance of these gases relative to that of their primary stars. In the solar system, Jupiter is enhanced by a factor of about 4 and Saturn a factor of 10 in C for example. The specific fingerprint of gas abundances is understood to provide insight into the reservoir of material that impacted the growing planets during their formation (Fortney et al. 2013; Lunine et al. 2019). Early trends in transiting planets show some evidence that the metallicity enhancement may be inversely proportional to planet mass. Whether such trends continue for colder giants that potentially formed farther from their stars and did not migrate is of particular interest.

The major species detected in reflected light spectra of the gas giant planets will almost certainly be those predicted by chemical equilibrium, as their atmospheres are dense and well mixed (trace amounts of disequilibrium species, such as CO or $PH_3$, are unlikely to be detectable for the cool giants that will be characterized by direct imaging). Thus optical to near-IR spectroscopy has the potential to measure true C, N, and O abundances. Figure 2 shows some of the diversity of reflected light spectra expected from various cool giant planets. Methane abundance and cloud top pressure can be derived with appropriate measurements of such spectra ($R > 50$, SNR > 15, e.g., Lupu et al. 2016, Batalha et al. 2019). Tests of abundance retrieval techniques on the relatively well understood giant planets will also give confidence that such methods are valid when applied to potentially habitable terrestrial exoplanet atmospheres characterized by the same instruments (e.g., Feng et al. 2018).

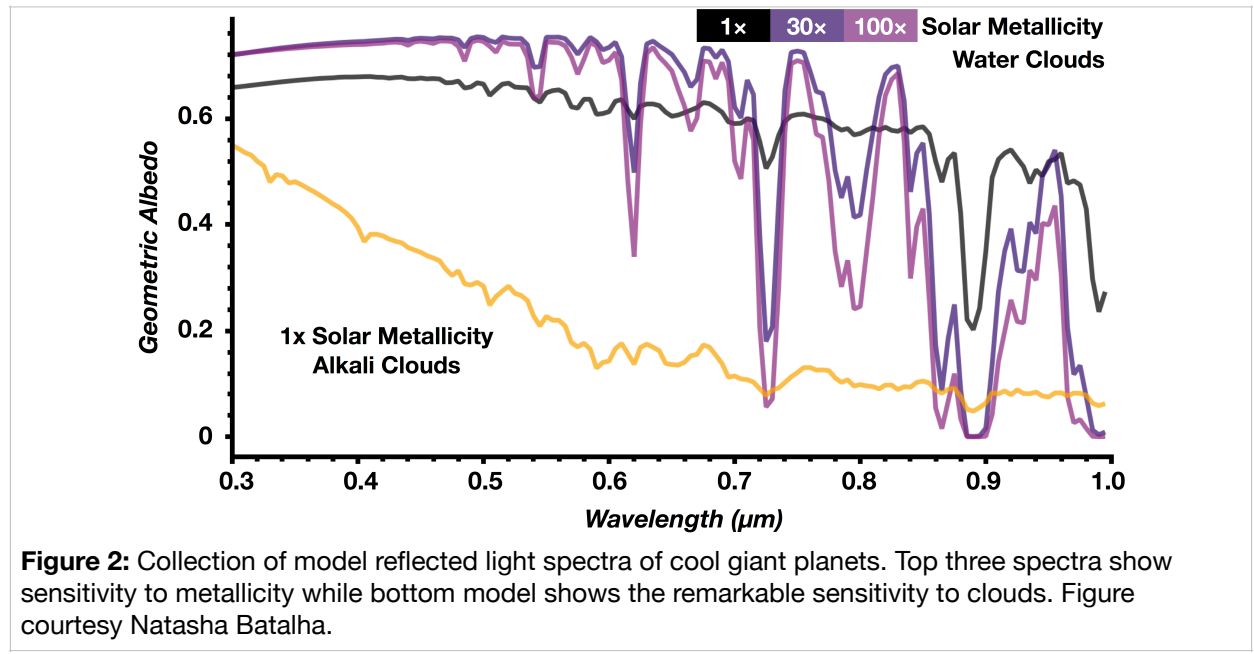

**Figure 2:** Collection of model reflected light spectra of cool giant planets. Top three spectra show sensitivity to metallicity while bottom model shows the remarkable sensitivity to clouds. Figure courtesy Natasha Batalha.

*Clouds*

Clouds and hazes in planetary atmospheres both inside and outside of the solar system critically shape reflected-light spectra. The influence of clouds on planetary spectra is notoriously difficult to predict, and clouds are a leading source of uncertainty in terrestrial climate models. With their onion-skin-like atmospheric structure, giant exoplanets offer a continuous array of cloud types to scatter incident flux. Figure 2 highlights the impact of $H_2O$ clouds on the reflected-light spectrum of a giant planet somewhat warmer than Jupiter. In fact, giant exoplanets in the habitable zones well outside the tidal locking distance of their host stars will provide some of our first opportunities to understand the formation of water clouds and their effect on global energy budgets in exoplanet atmospheres. The robustness with which we are able to constrain the properties of clouds (e.g., cloud top pressures, degree of partial cloudiness, single scattering albedos, and scattering asymmetry factors) will critically impact our ability understand clouds and extract the planetary thermal and chemical structure of both terrestrial and giant planets.

*Photochemistry*

Photochemical processes also play key roles in shaping the atmospheres and observed spectra of all solar system planets. In the stratospheres of solar system giant planets, $CH_4$ photochemistry generates hydrocarbons such as $C_2H_6$ and $C_2H_4$, and these can polymerize into more complex hydrocarbon species, some of which form aerosols. These hydrocarbons strongly absorb UV and blue light on Jupiter and Saturn. Hydrocarbon hazes are also crucial components of the atmosphere of Titan and likely were important at some periods in the atmosphere of the Archean Earth. In addition, giant exoplanets somewhat warmer than Jupiter will likely host an array of photochemical S, N, and O species. Lab results point to the likely ubiquity of photochemical hazes under diverse conditions (Hörst et al. 2018). Any complete characterization of an exoplanet atmosphere should account for the presence of hazes. Studies of hazy giant planets in reflected light would provide a valuable proving ground for understanding photochemical processes in atmospheres that differ from those found in the solar system and would give insight into habitable exoplanet haze processes.

Understanding the diversity of photochemical outcomes for all types of planets is a challenge. Because they form at low stratospheric pressures and are small, haze particles are most apparent from their interaction with blue and UV light. Investigating how hazes relate to atmospheric thermochemical structure, and surface composition/processes for terrestrial planets, will be a complex task, but one that will greatly inform our understanding of such processes as a whole.

**Conclusions**

The path forward for characterization of potentially habitable worlds through direct imaging proposed here mimics in many ways the path that has developed in the study of transiting exoplanets. Observational strategies, data reduction and interpretation techniques, and atmospheric modeling efforts were greatly refined on a sizeable (>100) population of giant exoplanets before inroads were made for transiting terrestrial exoplanets. It is our hope that any future facilities that aim to study exoplanets in detail acknowledge the criticality of exploring and learning from giant planets along the path towards characterizing habitable terrestrial planets.